\documentstyle[12pt]{article} 
\newcommand{\nc}{\newcommand}
\nc{\renc}{\renewcommand}

\nc{\half}{{\textstyle{1\over2}}}

\nc{\etal}{\mbox{\it et al. }}
\nc{\ie}{{\it i.e.}}
\nc{\eg}{{\it e.g.}}

\renc{\thefootnote}{\arabic{footnote}}
\nc{\capt}[1]{{\bf Figure.} {\small\sl #1}}

\nc{\eqs}[2]{\mbox{Eqs.~(\ref{#1},\,\ref{#2})}}
\nc{\eq}[1]{\mbox{Eq.~(\ref{#1})}}

\nc{\figs}[2]{\mbox{Figs.~(\ref{#1},\,\ref{#2})}}
\nc{\fig}[1]{\mbox{Fig~.(\ref{#1})}}

\nc{\tag}[1]{\label{#1} \marginpar{{\footnotesize #1}}}
\nc{\mtag}[1]{\label{#1} \mbox{\marginpar{{\footnotesize #1}}}}
\renc{\baselinestretch}{1.2}
\newlength{\overeqskip}
\newlength{\undereqskip}
\nc{\be}[1]{\begin{equation} \mbox{$\label{#1}$}}
\nc{\bea}[1]{\begin{eqnarray} \mbox{$\label{#1}$}}
\nc{\Section}[2]{\section{#2}\label{#1}}
\nc{\Bibitem}[1]{\bibitem{#1}}
\nc{\Label}[1]{\label{#1}}
\nc{\eea}{\vspace{\undereqskip}\end{eqnarray}}
\nc{\ee}{\vspace{\undereqskip}\end{equation}}
\nc{\bdm}{\begin{displaymath}}
\nc{\edm}{\end{displaymath}}
\nc{\dpsty}{\displaystyle}
\nc{\bc}{\begin{center}}
\nc{\ec}{\end{center}}
\nc{\ba}{\begin{array}}
\nc{\ea}{\end{array}}
\nc{\bab}{\begin{abstract}}
\nc{\eab}{\end{abstract}}
\nc{\btab}{\begin{tabular}}
\nc{\etab}{\end{tabular}}
\nc{\bit}{\begin{itemize}}
\nc{\eit}{\end{itemize}}
\nc{\ben}{\begin{enumerate}}
\nc{\een}{\end{enumerate}}
\nc{\bfig}{\begin{figure}}
\nc{\efig}{\end{figure}}
\nc{\arreq}{&\!=\!&}
\nc{\arrmi}{&\!-\!&}
\nc{\arrpl}{&\!+\!&}
\nc{\arrap}{&\!\!\!\approx\!\!\!&}
\nc{\non}{\nonumber\\*}
\nc{\align}{\!\!\!\!\!\!\!\!&&}

\def\lsim{\; \raise0.3ex\hbox{$<$\kern-0.75em
      \raise-1.1ex\hbox{$\sim$}}\; }
\def\gsim{\; \raise0.3ex\hbox{$>$\kern-0.75em
      \raise-1.1ex\hbox{$\sim$}}\; }
\nc{\DOT}{\hspace{-0.08in}{\bf .}\hspace{0.1in}}
\nc{\Laada}{\hbox {$\sqcap$ \kern -1em $\sqcup$}}
\nc\loota{{\scriptstyle\sqcap\kern-0.55em\hbox{$\scriptstyle\sqcup$}}}
\nc\Loota{{\sqcap\kern-0.65em\hbox{$\sqcup$}}}
\nc\laada{\Loota}
\nc{\qed}{\hskip 3em \hbox{\BOX} \vskip 2ex}

\nc{\real}{{\rm I \! R}}
\nc{\Z}{{\sf Z \!\!\! Z}}
\nc{\complex}{{\rm C\!\!\! {\sf I}\,\,}}
\def\bigid{\leavevmode\hbox{\small1\kern-3.8pt\normalsize1}}
\def\id{\leavevmode\hbox{\small1\kern-3.3pt\normalsize1}}
\nc{\slask}{\!\!\!/}
\nc{\bis}{{\prime\prime}}
\nc{\pa}{\partial}
\nc{\na}{\nabla}
\nc{\ra}{\rangle}
\nc{\la}{\langle}
\nc{\goto}{\rightarrow}
\nc{\swap}{\leftrightarrow}

\nc{\EE}[1]{ \mbox{$\cdot10^{#1}$} }
\nc{\abs}[1]{\left|#1\right|}
\nc{\at}[2]{\left.#1\right|_{#2}}
\nc{\norm}[1]{\|#1\|}
\nc{\abscut}[2]{\Abs{#1}_{\scriptscriptstyle#2}}
\nc{\vek}[1]{{\rm\bf #1}}
\nc{\integral}[2]{\int\limits_{#1}^{#2}}
\nc{\inv}[1]{\frac{1}{#1}}
\nc{\dd}[2]{{{\partial #1}\over{\partial #2}}}
\nc{\ddd}[2]{{{{\partial}^2 #1}\over{\partial {#2}^2}}}
\nc{\dddd}[3]{{{{\partial}^2 #1}\over
        {\partial #2 \partial #3}}}
\nc{\dder}[2]{{{d #1}\over{d #2}}}
\nc{\ddder}[2]{{{d^2 #1}\over{d {#2}^2}}}
\nc{\dddder}[3]{{d^2 #1}\over
        {d #2 d #3}}
\nc{\dx}[1]{d\,^{#1}x}
\nc{\dy}[1]{d\,^{#1}y}
\nc{\dz}[1]{d\,^{#1}z}
\nc{\dl}[1]{\frac{d\,^{#1}l}{(2\pi)^{#1}}}
\nc{\dk}[1]{\frac{d\,^{#1}k}{(2\pi)^{#1}}}
\nc{\dq}[1]{\frac{d\,^{#1}q}{(2\pi)^{#1}}}

\nc{\cc}{\mbox{$c.c.$ }}
\nc{\hc}{\mbox{$h.c.$ }}
\nc{\cf}{cf.\ }
\nc{\erfc}{{\rm erfc}}
\nc{\Tr}{{\rm Tr\,}}
\nc{\tr}{{\rm tr\,}}
\nc{\pol}{{\rm pol}}
\nc{\sign}{{\rm sign}}
\nc{\bfT}{{\bf T }}

\nc{\cA}{{\cal A}}
\nc{\cB}{{\cal B}}
\nc{\cD}{{\cal D}}
\nc{\cE}{{\cal E}}
\nc{\cG}{{\cal G}}
\nc{\cH}{{\cal H}}
\nc{\cL}{{\cal L}}
\nc{\cO}{{\cal O}}
\nc{\cT}{{\cal T}}
\nc{\cN}{{\cal N}}
\nc{\rvac}[1]{|{\cal O}#1\rangle}
\nc{\lvac}[1]{\langle{\cal O}#1|}
\nc{\rvacb}[1]{|{\cal O}_\beta #1\rangle}
\nc{\lvacb}[1]{\langle{\cal O}_\beta #1 |}
\nc{\bb}{\bar{\beta}}
\nc{\bt}{\tilde{\beta}}
\nc{\ctH}{\tilde{\cal H}}
\nc{\chH}{\hat{\cal H}}
\nc{\1}{\aa}
\nc{\2}{\"{a}}
\nc{\3}{\"{o}}
\nc{\4}{\AA}
\nc{\5}{\"{A}}
\nc{\6}{\"{O}}
\nc{\al}{\alpha}
\nc{\g}{\gamma}
\nc{\Del}{\Delta}
\nc{\e}{\epsilon}
\nc{\eps}{\epsilon}
\nc{\lam}{\lambda}
\nc{\om}{\omega}
\nc{\Om}{\Omega}
\nc{\ve}{\varepsilon}
\nc{\mn}{{\mu\nu}}
\nc{\k}{\kappa}
\nc{\vp}{\varphi}

\nc{\advp}[3]{{\it  Adv.\ in\ Phys.\ }{{\bf #1} {(#2)} {#3}}}
\nc{\annp}[3]{{\it  Ann.\ Phys.\ (N.Y.)\ }{{\bf #1} {(#2)} {#3}}}
\nc{\apl}[3]{{\it  Appl. Phys. Lett. }{{\bf #1} {(#2)} {#3}}}
\nc{\apj}[3]{{\it  Ap.\ J.\ }{{\bf #1} {(#2)} {#3}}}
\nc{\apjl}[3]{{\it  Ap.\ J.\ Lett.\ }{{\bf #1} {(#2)} {#3}}}
\nc{\app}[3]{{\it Astropart.\ Phys.\ }{{\bf #1} {(#2)} {#3}}}
\nc{\cmp}[3]{{\it  Comm.\ Math.\ Phys.\ }{{ \bf #1} {(#2)} {#3}}}
\nc{\cqg}[3]{{\it  Class.\ Quant.\ Grav.\ }{{\bf #1} {(#2)} {#3}}}
\nc{\epl}[3]{{\it  Europhys.\ Lett.\ }{{\bf #1} {(#2)} {#3}}}
\nc{\ijmp}[3]{{\it Int.\ J.\ Mod.\ Phys.\ }{{\bf #1} {(#2)} {#3}}}
\nc{\ijtp}[3]{{\it Int.\ J.\ Theor.\ Phys.\ }{{\bf #1} {(#2)} {#3}}}
\nc{\jmp}[3]{{\it  J.\ Math.\ Phys.\ }{{ \bf #1} {(#2)} {#3}}}
\nc{\jpa}[3]{{\it  J.\ Phys.\ A\ }{{\bf #1} {(#2)} {#3}}}
\nc{\jpc}[3]{{\it  J.\ Phys.\ C\ }{{\bf #1} {(#2)} {#3}}}
\nc{\jap}[3]{{\it J.\ Appl.\ Phys.\ }{{\bf #1} {(#2)} {#3}}}
\nc{\jpsj}[3]{{\it J.\ Phys.\ Soc.\ Japan\ }{{\bf #1} {(#2)} {#3}}}
\nc{\lmp}[3]{{\it Lett.\ Math.\ Phys.\ }{{\bf #1} {(#2)} {#3}}}
\nc{\mpl}[3]{{\it  Mod.\ Phys.\ Lett.\ }{{\bf #1} {(#2)} {#3}}}
\nc{\ncim}[3]{{\it  Nuov.\ Cim.\ }{{\bf #1} {(#2)} {#3}}}
\nc{\np}[3]{{\it  Nucl.\ Phys.\ }{{\bf #1} {(#2)} {#3}}}
\nc{\pr}[3]{{\it Phys.\ Rev.\ }{{\bf #1} {(#2)} {#3}}}
\nc{\pra}[3]{{\it  Phys.\ Rev.\ A\ }{{\bf #1} {(#2)} {#3}}}
\nc{\prb}[3]{{\it  Phys.\ Rev.\ B\ }{{{\bf #1} {(#2)} {#3}}}}
\nc{\prc}[3]{{\it  Phys.\ Rev.\ C\ }{{\bf #1} {(#2)} {#3}}}
\nc{\prd}[3]{{\it  Phys.\ Rev.\ D\ }{{\bf #1} {(#2)} {#3}}}
\nc{\prl}[3]{{\it Phys.\ Rev.\ Lett.\ }{{\bf #1} {(#2)} {#3}}}
\nc{\pl}[3]{{\it  Phys.\ Lett.\ }{{\bf #1} {(#2)} {#3}}}
\nc{\prep}[3]{{\it Phys\. Rep.\ }{{\bf #1} {(#2)} {#3}}}
\nc{\prsl}[3]{{\it Proc.\ R.\ Soc.\ London\ }{{\bf #1} {(#2)} {#3}}}
\nc{\ptp}[3]{{\it  Prog.\ Theor.\ Phys.\ }{{\bf #1} {(#2)} {#3}}}
\nc{\ptps}[3]{{\it  Prog\ Theor.\ Phys.\ suppl.\ }{{\bf #1} {(#2)} {#3}}}
\nc{\physa}[3]{{\it  Physica\ A\ }{{\bf #1} {(#2)} {#3}}}
\nc{\physb}[3]{{\it  Physica\ B\ }{{\bf #1} {(#2)} {#3}}}
\nc{\phys}[3]{{\it Physica\ }{{\bf #1} {(#2)} {#3}}}
\nc{\rmp}[3]{{\it  Rev.\ Mod.\ Phys.\ }{{\bf #1} {(#2)} {#3}}}
\nc{\rpp}[3]{{\it Rep.\ Prog.\ Phys.\ }{{\bf #1} {(#2)} {#3}}}
\nc{\sjnp}[3]{{\it Sov.\ J.\ Nucl.\ Phys.\ }{{\bf #1} {(#2)} {#3}}}
\nc{\spjetp}[3]{{\it Sov.\ Phys.\ JETP\ }{{\bf #1} {(#2)} {#3}}}
\nc{\yf}[3]{{\it Yad.\ Fiz.\ }{{\bf #1} {(#2)} {#3}}}
\nc{\zetp}[3]{{\it Zh.\ Eksp.\ Teor.\ Fiz.\  }{{\bf #1}  {(#2)} {#3}}}
\nc{\zp}[3]{{\it Z.\ Phys.\ }{{\bf #1} {(#2)} {#3}}}
\nc{\ibid}[3]{{\sl ibid.\ }{{\bf #1} {#2} {#3}}}
\nc{\rf}[1]{(\ref{#1})}
\nc{\nn}{\nonumber \\*}
\nc{\bfB}{\bf{B}}
\nc{\bfv}{\bf{v}}
\nc{\bfx}{\bf{x}}
\nc{\bfy}{\bf{y}}
\nc{\vx}{\vec{x}}
\nc{\vy}{\vec{y}}
\nc{\oB}{\overline{B}}
\nc{\oI}{\overline{I}}
\nc{\oR}{\overline{R}}
\nc{\rar}{\rightarrow}
\nc{\ti}{\times}
\nc{\slsh}{\hskip-5pt/}
\nc{\sm}{Standard~Model~}
\nc{\MP}{M_{\rm Pl}}
\nc{\tp}{t_{\rm Pl}}
\nc{\ave}{\bar{E}}

\def\sepand{\rule{14cm}{0pt}\and}
\nc{\eff}{{\rm eff}}
\nc{\kk}{\vek{k}}
\nc{\pp}{{\rm p}}
\nc{\ga}{g_{a\gamma}}
\nc{\vv}{\\}
\nc{\eee}{{\bf E}}
\nc{\bbb}{{\bf B}}
\nc{\qcd}{T_{\rm QCD}}
\nc{\G}{\rm \ G}
\def\vec#1{{\bf #1}}
\begin{document}
\topmargin 0cm
\topskip 0mm   
 {\title
{\null\vskip-3truecm
{ \hskip10truecm {\small HU-TFT-96-35}\vskip 0.1cm}
%{ \hskip10truecm {\small hep-ph/9405316\hfill }\vskip 1.5cm}
{\bf The effect of Silk damping on primordial magnetic fields}}
\author{
{\sc Axel Brandenburg$^1$} \\
{\sl Department of Mathematics and Statistics},\\
{\sl University of Newcastle upon Tyne, NE1 7RU, UK}\\
and\\
{\sc Kari Enqvist$^2$} \\
{\sl Department of Physics} \\ 
{\sl University of Helsinki, FIN-00014 Helsinki, Finland}\\
and\\
{\sc Poul Olesen$^3$ }\\ 
{\sl The Niels Bohr Institute, University of Copenhagen,} \\
{\sl Blegdamsvej 17, DK-2100 Copenhagen, Denmark} \\
\sepand
}\maketitle}

\vspace{1.5cm}
\begin{abstract}
We study the effects of plasma viscosity on the dynamics of 
primordial magnetic fields by simulating magnetohydrodynamics in
the early universe by appropriate non-linear cascade models. We find 
numerically that even in the presence of large kinetic 
viscosity, magnetic energy is transferred to 
large length scales. There are indications,  however, that the
inverse cascade stops at a given time which depends on the 
magnitude of viscosity.
For realistic viscosities we do not find
equipartition between magnetic and kinetic energies.
\noindent
\end{abstract}

\vfil
\footnoterule
{\small  $^1$Axel.Brandenburg@Newcastle.ac.uk; $^2$enqvist@pcu.helsinki.fi;  
$^3$polesen@nbi.dk}
\thispagestyle{empty}
\newpage
\setcounter{page}{1}
%%%%%%%%%%%%%%%%%%%%%%%%%%%%%%%%%%%%%%%%%%%%%%%%%%%%%%%%%%%%%
%%%%%%%%%%%%%%%%%%%%%%%%%%%%%%%%%%%%%%%%%%%%%%%%%%%%%%%%%%%%%%%%%%%
There are many ways that magnetic fields could have been generated
at various stages in the early universe. In the theory of galactic
magnetism such fields are often referred to as ``cosmological'' or
``primordial'', because they are created well before galaxies are
formed. These fields may play the role of an initial condition
for the galactic dynamo, a mechanism that would amplify magnetic fields
and convert kinetic energy into magnetic \cite{bbmss96}.

There are two major problems when invoking primordial magnetic fields as
possible seed magnetic fields for the galactic dynamo. One problem is
the very small length scale of such magnetic fields. The horizon
scale at the time of the electroweak phase transition is just a few
centimeters, corresponding to about 1 AU at the present time. This is
nine orders of magnitude shorter than the radius of typical galaxies.
However, this view is too simplistic, because nonlinear effects 
inherent in the magnetohydrodynamic (MHD) equations can lead
to a redistribution of magnetic energy over different length scales.
Unfortunately, MHD in 3+1
exceeds the possibilities of present day computers.
Therefore one has to resort to models which simulate the
MHD equations. In a previous paper \cite{brandenburgetal}, 
hereafter referred to as Paper~1, we
developed a fully relativistic 3+1 d version of the 
so-called cascade model \cite{goy} appropriate for MHD and found that
an inverse cascade is operative, whereby magnetic energy is continuously 
tranferred to larger length scales.
The other problem is that around the time of recombination photon
diffusion becomes very large and could smooth out all fluctuations \cite{silk}.
This may then also destroy the magnetic field \cite{chicago}.
The purpose of this paper is to show that this too is too simplistic
a viewpoint, and that nonlinear effects most likely 
prevent this from happening.

The basic equations have been presented and discussed in Paper~1.
We started out from the fully relativistic MHD equations 
in expanding (flat) space  and showed that all
the terms arising from the expansion can be removed by using rescaled
quantities and conformal time,
\be{conf}
t=\int dt_H/R(t_H),
\ee
where $t_H$ is the Hubble time and $R(t_H)$ is the expansion factor.
Starting from random initial conditions, we obtained turbulent velocity
and magnetic fields, very much like those in ordinary (nonrelativistic)
decaying hydromagnetic turbulence. In order to study the effects
of this kind of turbulence we adopted a simple cascade model that captures
the qualitative features related to turbulent energy spectra. Such models
\cite{goy} have been rather successful in predicting even subtle corrections to
Kolmogorov turbulence due to intermittency effects \cite{mogens}.
(For a recent review, see ref.\ \cite{mogensbook}.)
Our model is more general in that it includes the effects of magnetic 
fields. Unlike nonmagnetic turbulence, in the presence of magnetic fields
there is an inverse cascade of magnetic helicity, which leads to a
transport of magnetic energy to larger and larger scales \cite{pfl76}.
Using our cascade model  (also called shell model)
we found that the integral scale
\be{intscale}
l_0=\left.\int (2\pi/k)E_M(k)dk\right/\int E_M(k)dk,
\ee
where $E_M(k)$ is the magnetic energy spectrum,
increases with the Hubble time approximately like $t_H^{0.25}$.

In Paper~1 we considered the case where the kinematic viscosity $\nu$
was equal to the magnetic diffusivity $\eta$ (inverse electrical
conductivity) which, in turn, was assumed to be small (correponding to
a large magnetic Reynolds number). However, around the time of
recombination the photon mean free path $\lambda_\gamma$ became very
large and photon diffusion became very efficient in smoothing out
virtually all inhomogeneities of the photon-baryon plasma \cite{kt90}.
This process if often referred to as Silk damping, which corresponds to
a kinematic viscosity $\nu\simeq\lambda_\gamma$ (in natural units).
We have computed numerically the evolution of magnetic and kinetic
energy spectra in two completely different cascade models and studied
the effects of viscosity. Our results are presented in Figs. 1 and 2 and the
main point can be summarized as follows: in the cascade models
magnetic energy is transferred to large length scales even in the presence of 
large viscosity. It seems likely that the same is true also in full 
3+1 MHD.

Let us first use the cascade model of Paper~1 to investigate the effect of
very large values of $\nu$ on the magnetic field.
In this model velocity and magnetic fields are described by the
variables $v_n$ and $b_n$, representing the ``collective'' behavior
at wavenumber $k_n$.
In the original model, $k_n=r^n$ with $r=2$.
Below we shall also consider a continuous version of this model with
$r\rightarrow1$, which was orginally studied by Parisi \cite{parisi}
in a hydrodynamical context. 
We reiterate here the salient features of the model. We present the 
equations of motion in a slightly different form with the negative and
positive helicity states split, to facilitate comparision with 
the subsequent continuous model:
\bea{evolut1}
(dv_n^+/dt+\nu k_n^2 v_n^+)^*&=&ik_n 
(v_{n+1}^-v_{n+2}^++\frac{1-r}{r^2}v_{n-1}^-v_
{n+1}^--\frac{1}{r^3}v_{n-1}^-v_{n-2}^+-b_{n+1}^-b_{n+2}^+\nn  
&-&\frac{1-r}{r^2}b_{n-1}^-b_{n+1}^-+\frac{1}{r^3}b_{n-1}^-b_{n-2}^+),
\eea
and
\bea{evolut2}
(db_n^+/dt+\eta k_n^2 b_n^+)^*&=&i\frac{k_n}{r(1+r)}~(v_{n+1}^-b_{n+2}^+-
b_{n+1}^-v_{n+2}^++v_{n-1}^-b_{n+1}^--b_{n-1}^-v_{n+1}^-\nn &-&
v_{n-1}^-b_{n-2}^++b_{n-1}^-v_{n-2}^+),
\eea
with similar equations with $v_n^+\leftrightarrow v_n^-$ and $b_n^+
\leftrightarrow b_n^-$. The model describes an expanding radiation
dominated magnetic universe $(p=\rho/3$). In Eqs. 
\rf{evolut1}-\rf{evolut2} $t$ is the conformal time, and the units are
such that $\rho\sim T^4$ (for details, see
\cite{brandenburgetal}).
Eqs.~\rf{evolut1}-\rf{evolut2} conserve energy and helicity in the 
absence of magnetic diffusion $\eta$ and plasma viscosity $\nu$, 
and + and -- are related to magnetic helicity. 
Thus, if $\nu=\eta=0$, the quantities
\be{cons}
E=\sum (|b_n^+|^2+|b_n^-|^2+|u_n^+|^2+|u_n^-|^2),~~
H=\sum (|b_n^+|^2-|b_n^-|^2)/k_n,
\ee
are conserved. This is a direct consequence of Eqs.~\rf{evolut1}-\rf{evolut2}.
In the case of ordinary MHD $b$ is just the magnetic field.
In a cosmological setting
with a flat universe, $b$ is the magnetic field multiplied by $R(t)^2$, 
and $t$ is conformal time. The conservation of helicity exhibited 
above is nontrivial in three, but not in two dimensions, so that 
in this sense the
cascade model is three dimensional.

There does not exist any proof that the cascade model and the standard MHD 
equations are equivalent.
There are features which are similar in both cases: the equations couple many
different scales (making it hard to predict anything a priori,
especially when the Reynolds number
is large and nonlinear effects are important), they have similar 
conservation laws, 
and the equations of motion are similar. Also, in the pure hydrodynamic 
case the cascade model
equations have been compared to experiments \cite{mogens} and good results on
intermittency have been obtained.  In
any case, the properties of turbulence have not been derived from first 
principles,
and the cascade model is therefore an interesting (toy?) tool.

For large diffusion coefficients the equations become stiff
and it is therefore essential to solve for the diffusion term
exactly. Using the identity
\begin{equation}
{dv_n\over dt}+\nu k_n^2 v_n=e^{-\nu k_n^2t}
{d\over dt}\left(v_n e^{\nu k_n^2t}\right)
\end{equation}
we solve equations of the form
\begin{equation}
{dv_n\over dt}+\nu k_n^2 v_n=N_n(t)
\end{equation}
by a modified second order Adams-Bashforth scheme
\begin{equation}
v_n(t+\delta t)=K_n\{v_n(t)+\textstyle{1\over2}\delta t
[3N_n(t)-K_n N(t-\delta t)]\}
\end{equation}
where
\begin{equation}
K_n=e^{-\nu\delta t k_n^2}.
\end{equation}

We start from an initial condition that yields a magnetic energy spectrum
similar to that found at later stages. Fig.~2 of Paper~1 suggests that
the spectrum has developed an inertial range which is approximately constant,
$E_M(k)={\rm const.}$ for $k_d<k<k_0$, where $k_0=2\pi/l_0$ is the wavenumber
corresponding to the integral scale and $k_d\gg k_0$ the wavenumber of the
diffusive cutoff scale. We adopt 30 wavenumber shells ($1<n<N$ with $N=30$)
and place the cutoff wavenumber at $n=27$ and use $\eta=10^{-11}$.
We put $v_n=0$ initially and compare the results for two different values
of $\nu$ ($10^{-2}$ and $10^2$). 

The resulting spectra are displayed in Fig.~1.
\begin{figure}
\leavevmode
\centering
\vspace*{90mm}
\includegraphics{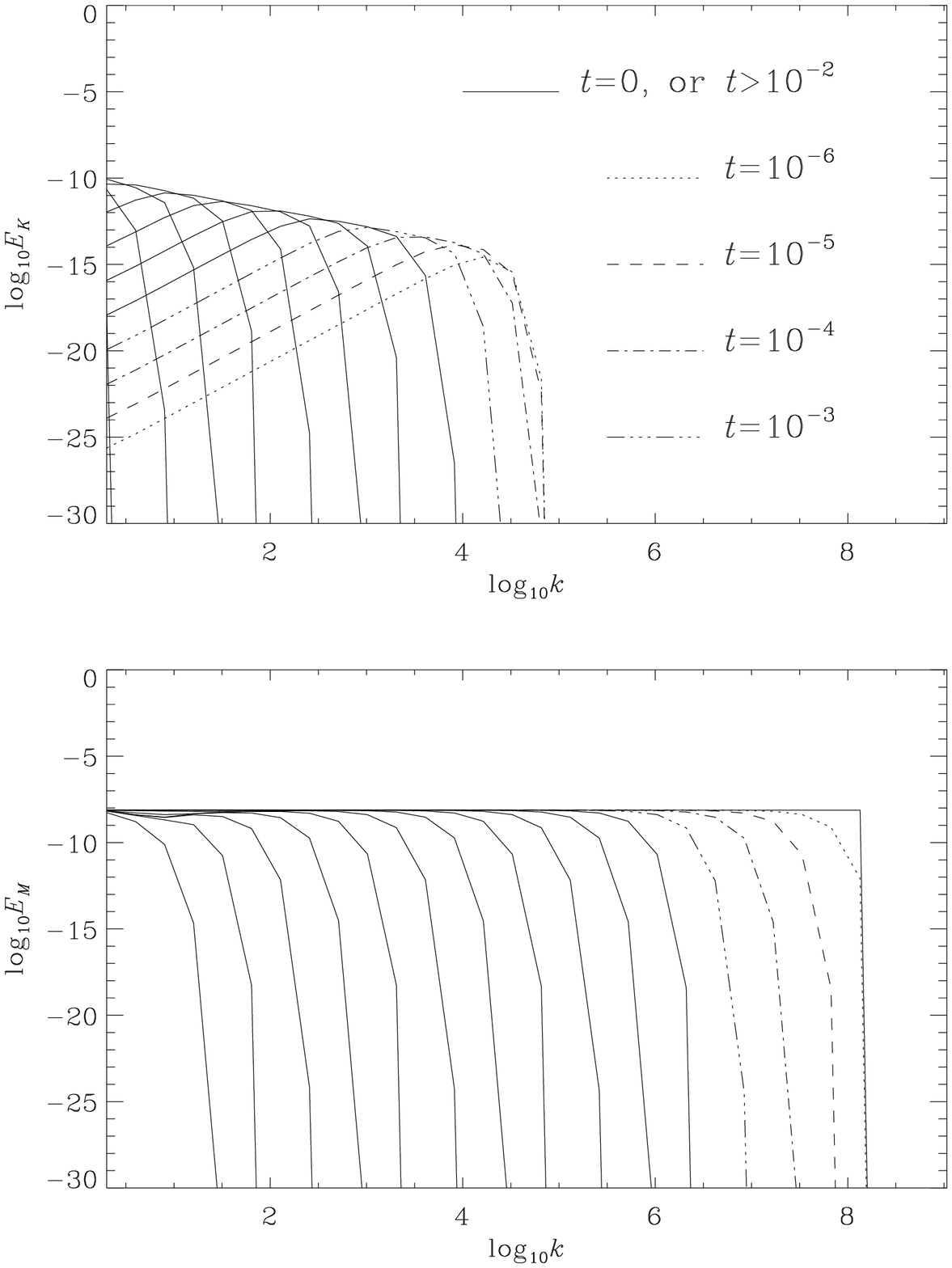}
\includegraphics{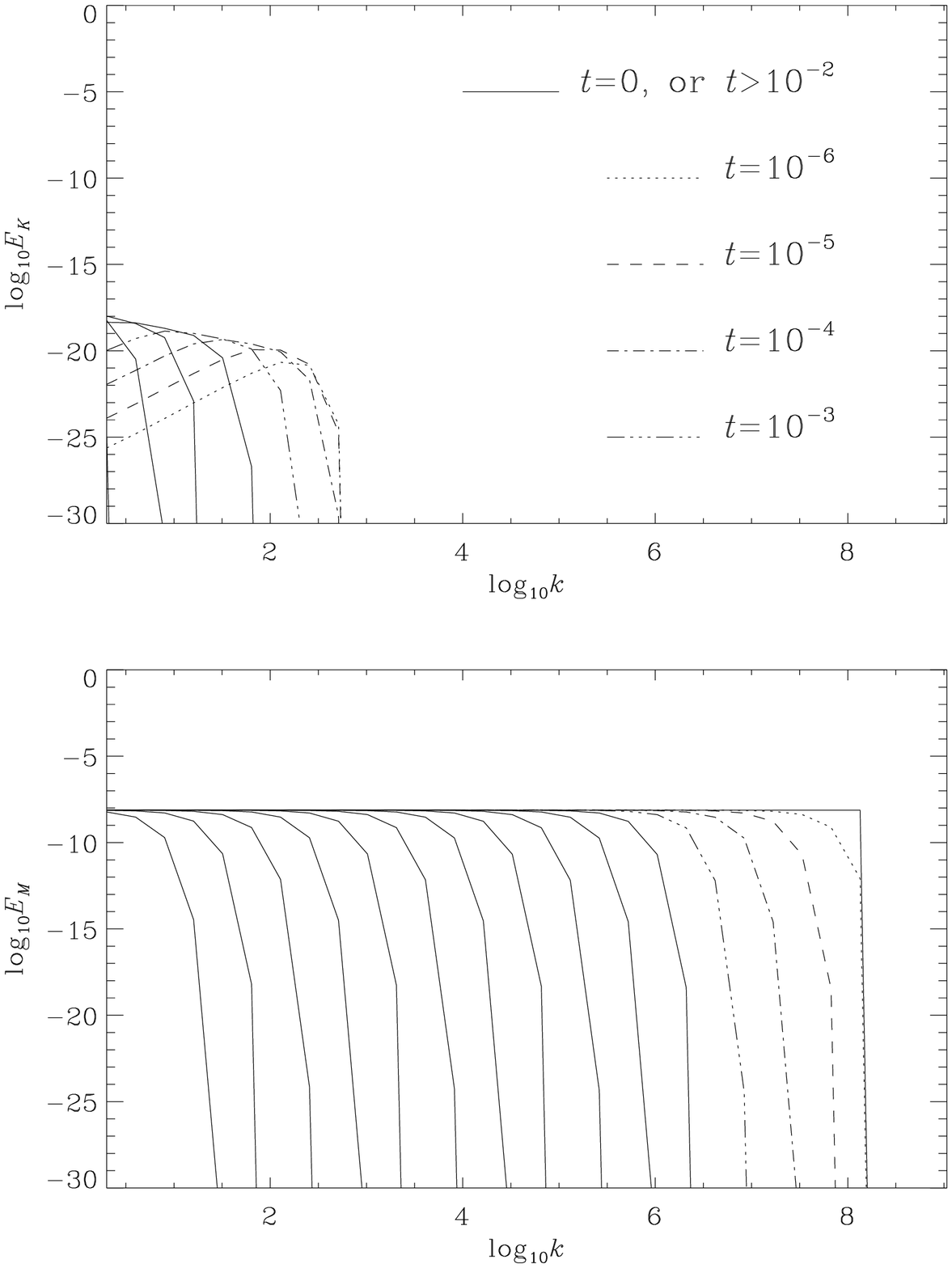}
\caption{Magnetic and plasma kinetic energy spectrum 
as a function of the wave number $k$ in the cascade model for  
small ($\nu=10^{-2}$) plasma viscosity (left) and large ($\nu=10^{2}$) 
plasma viscosity (right). The highest time is $t=10^{8}$, corresponding to
a Hubble time $10^{16}$}
\label{kuvaaxel}
\end{figure}
%\vspace*{30mm}
The results show that $v_n$ is generated from $b_n$, but it gets weaker if
$\nu$ is increased. At small $\nu$ and at large times
there is an approximate equipartition of the magnetic
and kinetic energies with $v_n=b_n=B_nR^2$, where $B_n$ is the unscaled
magnetic field. Since we are using units in which
the initial effective energy density $\rho+p=4\rho/3=1$ 
it follows that $v_n$ is the Alfven
velocity $B_n/\sqrt{\frac 43\rho}$
of the equilibrium plasma. For large $\nu$ equipartition is lost,
which signals the breakdown of the perturbative approach. As we shall 
argue later, the large $\nu$ case is appropriate for the very early universe. 
More importantly, Fig.~1 shows that the evolution of $b_n$ is not affected
by $v_n$ and $\nu$, but is rather governed by ohmic decay,
$b_n\sim\exp(-\eta k_n^2 t)$. It may also be seen that in the case of
large kinetic viscosity $\nu$=100 the velocity modes decay approximately 
according to $v_n\sim \exp(-\nu k_n^2t)$.

We shall now study the effect of a large $\nu$ on $b$. We have done some 
more detailed calculations in order to investigate what is behind
the behaviour shown in Fig. 1. The results are shown in Fig. 2,
which shows that for a sufficiently large viscosity, the inverse cascade
stops. 
\begin{figure}
\leavevmode
\centering
\vspace*{90mm}
\includegraphics{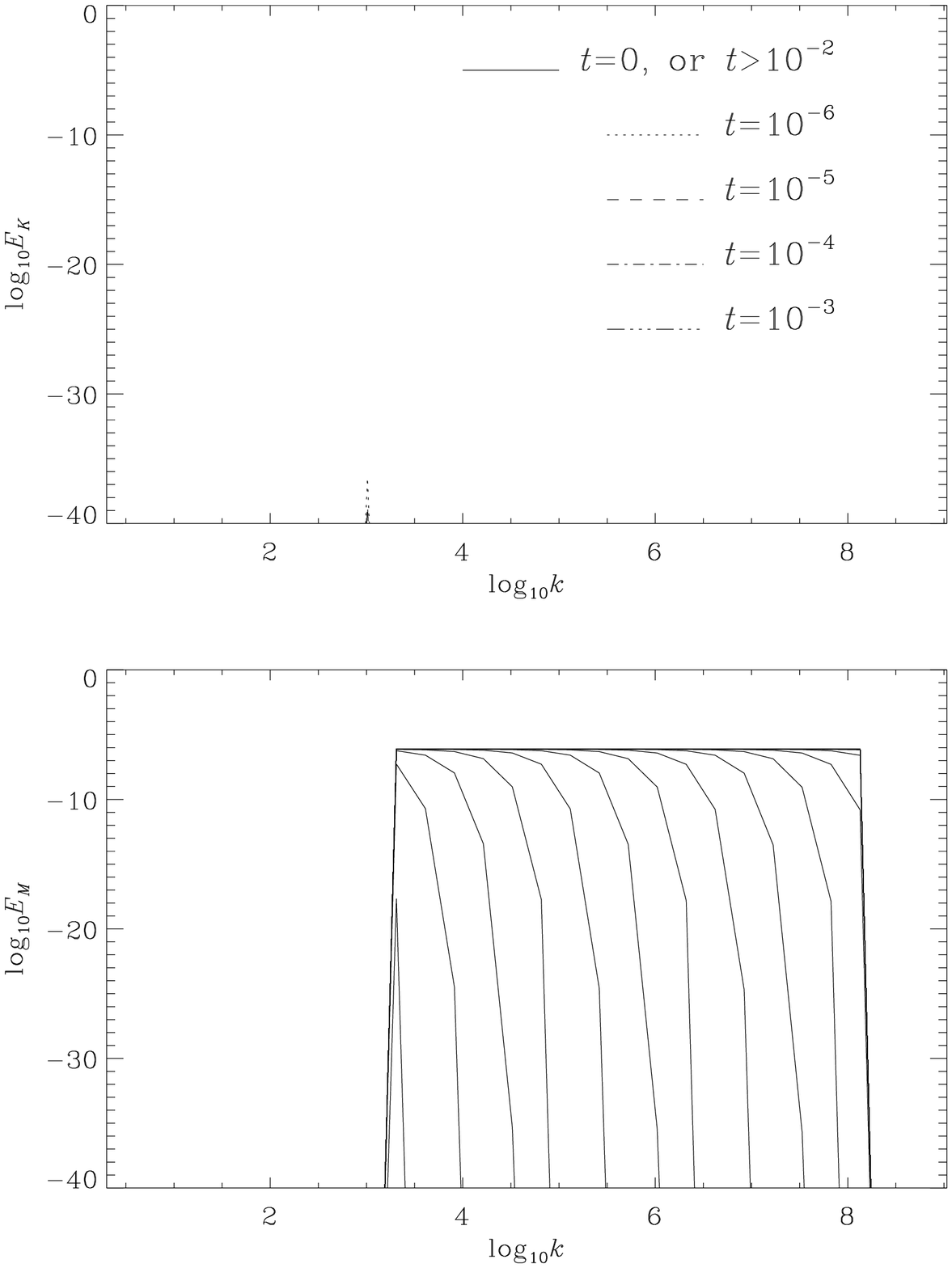}
\includegraphics{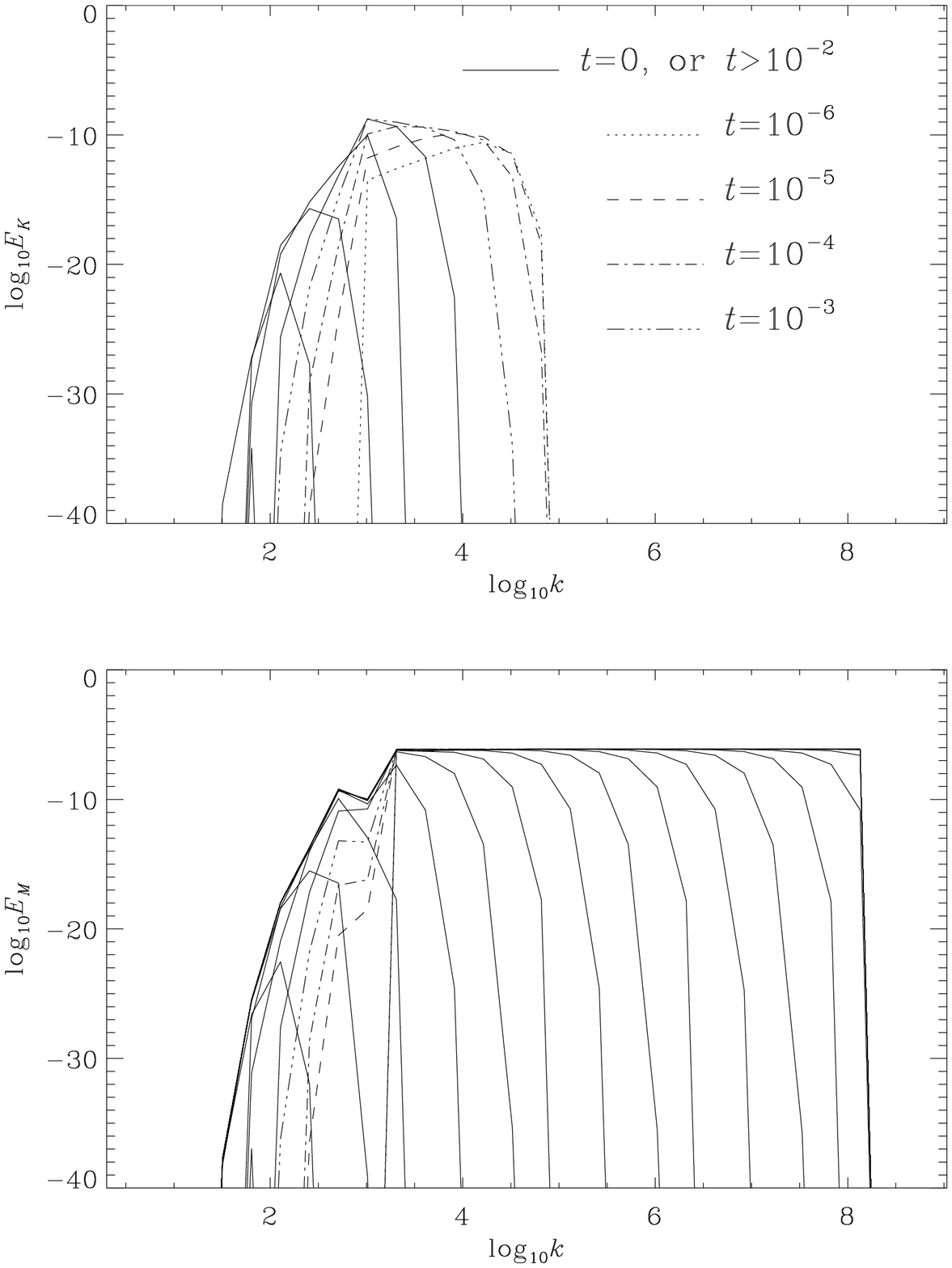}
\caption{The behaviour of the cascade model for $\nu=10^2$ 
(left) and $\nu=10^{-2}$ (right)
at large times (the last time slice is $t=10^{37}$).
The figure shows what happens to an initially flat spectrum (which
does not extend to k=0). 
Here $\eta=10^{-31}$.}
\label{kuvaaxel2}
\end{figure}
%\vspace*{30mm}
With a magnetic enegy of 10$^{-8}$, the cascade goes over five decades
in the most unfavourable case $\nu$=100. The inverse cascade stops at some 
low value of k, and the field just disappears by Ohmic decay.
One can then ask, at which temperature does the inverse 
cascade stop?
To estimate this, one studies  the ratio between the non-linear terms and 
the Silk viscosity term, i.e. the Reynolds number.
We start at the annihilation $t_H\sim 1$ sec, where we assume that there
are no velocity fluctuations. This is not true if the primordial field
originates at the electroweak phase transition $T\sim$ 100 GeV, since there
will always be velocity fluctuations associated with the magnetic field.
However, to allow for the optimal conditions for Silk diffusion, 
we assume that
there are no velocity fluctuations at the annihilation. 
We then have to estimate
the resulting Reynolds number,
\be{sveske}
{\rm Re}\sim\frac{ v}{k\nu_{Silk}/R},
\ee
where the viscosity  $\nu_{Silk}/R$ is effectively given by the photon
diffusion length, 
\be{kuk}
\nu\sim \frac{{\rm 10}^{6}}{T^3}~{\rm GeV}^2.
\ee 
As mentioned above, the inverse cascade spans over five decades in $k$. Thus, 
the typical $k$ which enters in $Re$ in eq. \rf{sveske} is given by
\be{hannibal}
k={\rm 10}^{-5}/x t_H^{ann},
\ee
where $xt_H^{ann}$ is the spatial scale scale of the magnetic field at
the annihilation. The fraction $x$ is thus determined by the mechanism
for producing the primordial field. Before annihilation, we assume that the
viscosity is so unimportant that it can be ignored. This is why we
 start at annihilation, where Silk diffusion becomes very
large. The inverse cascade stops when
$Re$ is of order one, i.e. when the diffusion equals the non-linear terms.
We may therefore conclude that in the particular case displayed in
Fig. 2 ($\nu$=100), with $v=10^{-4}$, the cascade stops close to
recombination.

To verify that the above results are generic and not just a feature of
the model adopted, we will now consider another model of the full 3+1
MHD \cite{parisi}. Formally it may be obtained from  \rf{evolut1}-\rf{evolut2}
by passing to the limit $r\rightarrow$1, but it
may also be viewed as
a completely independent model of MHD. Writing $r=1+\epsilon$ we get
\bea{cont1}         
\left({\pa v^+(k,t)\over\pa \epsilon t}+\nu k_n^2 v^+(k,t)\right)^* &=&
\hskip-2pt
ik\Bigl(4v^-k{\pa v^+\over\pa k}+2v^+k{\pa v^-\over\pa k}+3v^-v^+-(v^-)^2 \nn
&-&\hskip-2pt
4b^-k{\pa b^+\over\pa k}-2b^+k{\pa b^-\over\pa k}-3b^+b^-+(b^+)^2\Bigr)
+{\cal O}(\epsilon)
\eea
and
\bea{cont2}
\left({\pa b^+(k,t)\over\pa \epsilon t}+\eta k_n^2 b^+(k,t)\right)^*&=&
ik\Bigl(b^+k{\pa v^-\over\pa k}+2v^-k{\pa b^+\over\pa k}-v^
+k{\pa b^-\over\pa k}-2b^-k{\pa v^+\over\pa k} \nn
&+&v^-k{\pa b^-\over\pa k}-b^-k{\pa v^-\over\pa k}\Bigr)+{\cal O}(\epsilon).
\eea
In addition to eqs. \rf{cont1} and \rf{cont2} there are two equations more, 
obtained by 
making the replacements $b^+\leftrightarrow b^-$ and $v^+\leftrightarrow v^-$.
These equations
conserve the quantities
\be{ccons}
E=\int \frac{dk}{k} (|v^+|^2+|v^-|^2+|b^+|^2+|b^-|^2),~~
H=\int \frac{dk}{k^2} (|b^+|^2-|b^-|^2),
\ee
for $\nu=\eta=$0, provided the following boundary conditions are satisfied for 
$k\rightarrow$0
and $k\rightarrow \infty$,
\bea{bound1}
kv^-(v^+)^2&\rightarrow& 0,~kv^+b^+b^-\rightarrow 0,~kv^-b^+\rightarrow 0,\nn
kv^+(v^-)^2&\rightarrow& 0,kv^-b^+b^-\rightarrow 0,kv^+(b^-)^2\rightarrow 0,
\eea
and
\be{bound2}
v^-(b^+)^2\rightarrow 0,v^+(b^-)^2\rightarrow 0,b^+b^-v^+\rightarrow 0,
b^+b^-v^-\rightarrow 0,
\ee 
respectively. If the conditions at infinity are not satisfied, 
this corresponds to ``diffusion at infinity".

We shall now simplify the continuous cascade  model by using a scaling 
first introduced by
Parisi \cite{parisi} for the case of pure hydrodynamics without helicity 
conservation. If one
considers the original discrete equations \rf{evolut1}-\rf{evolut2}
in the absence of viscosity, and assume that we start at time 
$t$=0 with a {\it primordial} spectrum
\be{primordial}
b^+(k,0)=b^-(k,0)=k^p,~~v^+(k,0)=v^-(k,0)=0,
\ee
then it is easy to see that the discrete 
 cascade model  equations have
a solution with the scaling form
\bea{scaling}
b^+(k,t)&=&k^pB(k^{1+p}t),b^-(k,t)=k^pM(k^{1+p}t),\nn
v^+(k,t)&=&k^pu(k^{1+p}t),v^-(k,t)
=k^pm(k^{1+p}t).
\eea 
These equations imply that $E_B/k^{2p-1}$ is a function of 
$k^{p+1}t$ only.

If we insert the scaling \rf{scaling} in the continuous cascade  model Eqs. 
\rf{cont1}-\rf{cont2},
then we see that powers of $k$ cancel out neatly, and we are left with the 
coupled equations\footnote{We have here selected the phase of $b$ and $v$ to 
be -$\pi/2$. With a dynamical phase the number of equations double from 
four to eight, since each
$b$ and $v$ have a real and imaginary part.}
\bea{u}
-u'(x)&=&3(2p+1)um+4(1+p)xmu'+2(1+p)xum'-m^2
\nn &-&3(1+2p)MB-4(1+p)xMB'-2(1+p)xBM'+M^2,
\eea
and
\bea{B}
-B'(x)&=&3pmB-3puM+2(1+p)xmB'+(1+p)xBm'-(1+p)xuM'\nn 
&-&2(1+p)xMu'+(1+p)xmM'-(1+p)xMm',
\eea
as well as two equations where the interchanges $B\leftrightarrow M$ 
and $u\leftrightarrow m$ have
been made. Thus, the original set of $2N$ (with $n\leq 2N$)  coupled 
differential equations
have been replaced by only four. In these equations
\be{def}
x=k^{1+p}t
\ee
is the scaling variable.

It should be emphasized that the continuous cascade model
is a priori ``as good as" the discrete
one, since in both cases the conservation equations are 
satisfied when there is no
diffusion (except for possible diffusion at infinity). 
It may of course turn out that
phenomenologically one type is better than the other.

In Paper 1 we found that if one starts with a primordial 
spectrum $p=$3/2, then
there is an inverse cascade, transferring energy from large to small $k$. 
The scaling eqs. \rf{scaling} satisfy this: For a fixed value of the variable 
$x$ in \rf{def},
the functions $B,M,u,~{\rm and}~m$ have some definite values. Thus, as 
time is increased, these values remain the same if $k$ is 
diminished in such a way that  $x$ remains
constant. Thus, the scaling relations \rf{scaling} predicts an inverse cascade.
  
  This is also seen if we consider the integral scale (the ``correlation
length"), given by \eq{intscale},
which measures the large scale structures in comoving coordinates. We may 
consider $l_0$
as the expectation value of 1/$k$. From  scaling we therefore expect that
\be{behavior}
l_0~\approx t^{1/(1+p)}~\approx t_H^{1/2(1+p)},
\ee
where $k^p$ is the initial primordial spectrum. From this we see that for 
$p$=3/2 one obtains
\be{behavior2}
l_0\approx t_H^{0.2},
\ee
where the power is in essential agreement with the previously found 
\cite{brandenburgetal}
value $p\approx$ 0.25. 

Since we do not commit ourselves to any specific model, let us consider an 
initial spectrum
$k^p$. The integral scale then behaves as in \rf{behavior} in comoving 
coordinates, which
means that in physical coordinates it goes like
\be{physical}
l_0^{physical}~\approx t_H^{\frac{2+p}{2(1+p)}}.
\ee
Thus, if there exists a model with $p$=0, in such a case the large scale 
structures are of
the order the horizon. The case $p$=0 corresponds to a scale invariant 
primordial spectrum
$dk/k$.

\begin{figure}
\leavevmode
\centering
\vspace*{60mm}
\includegraphics{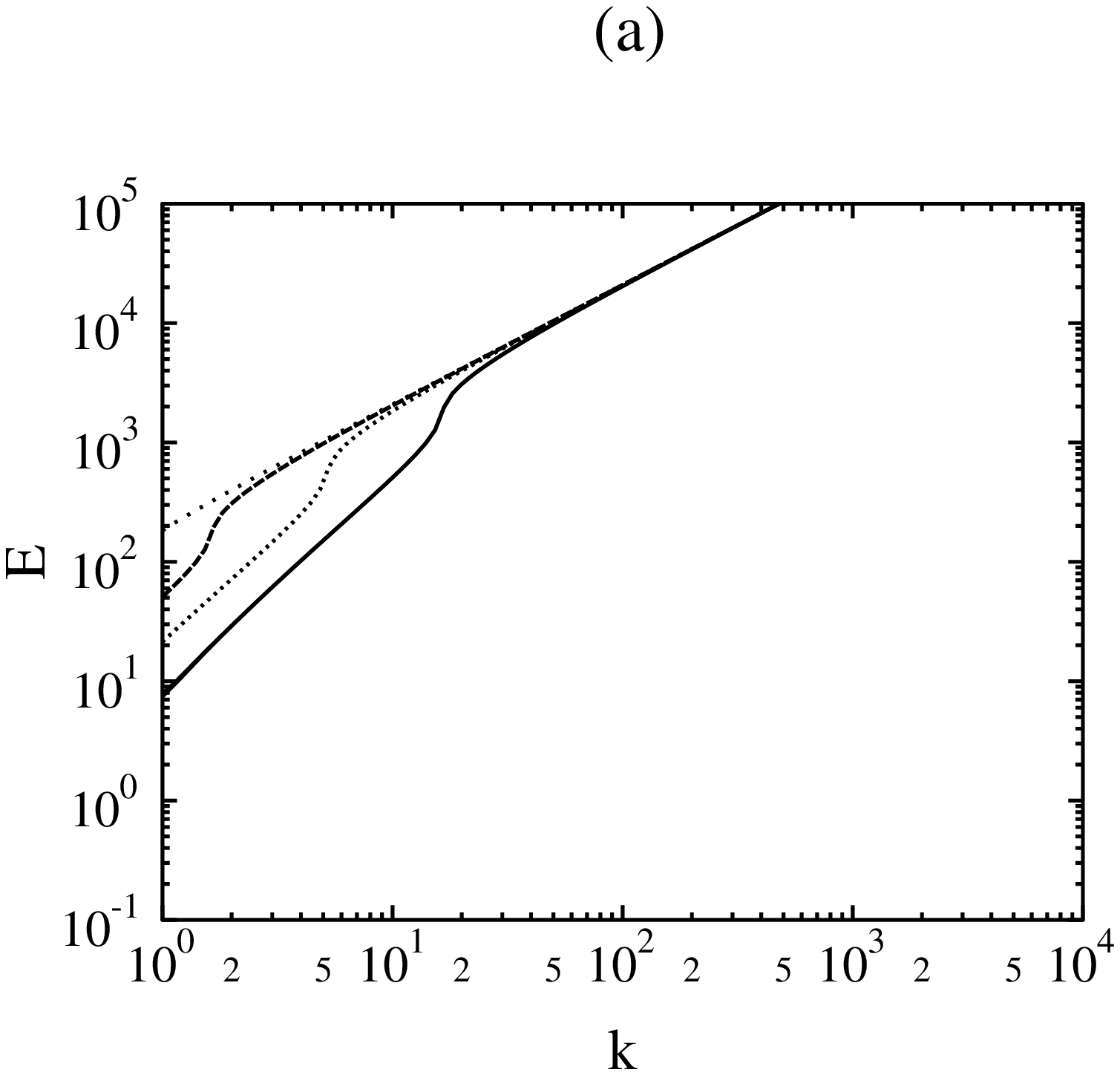}
\includegraphics{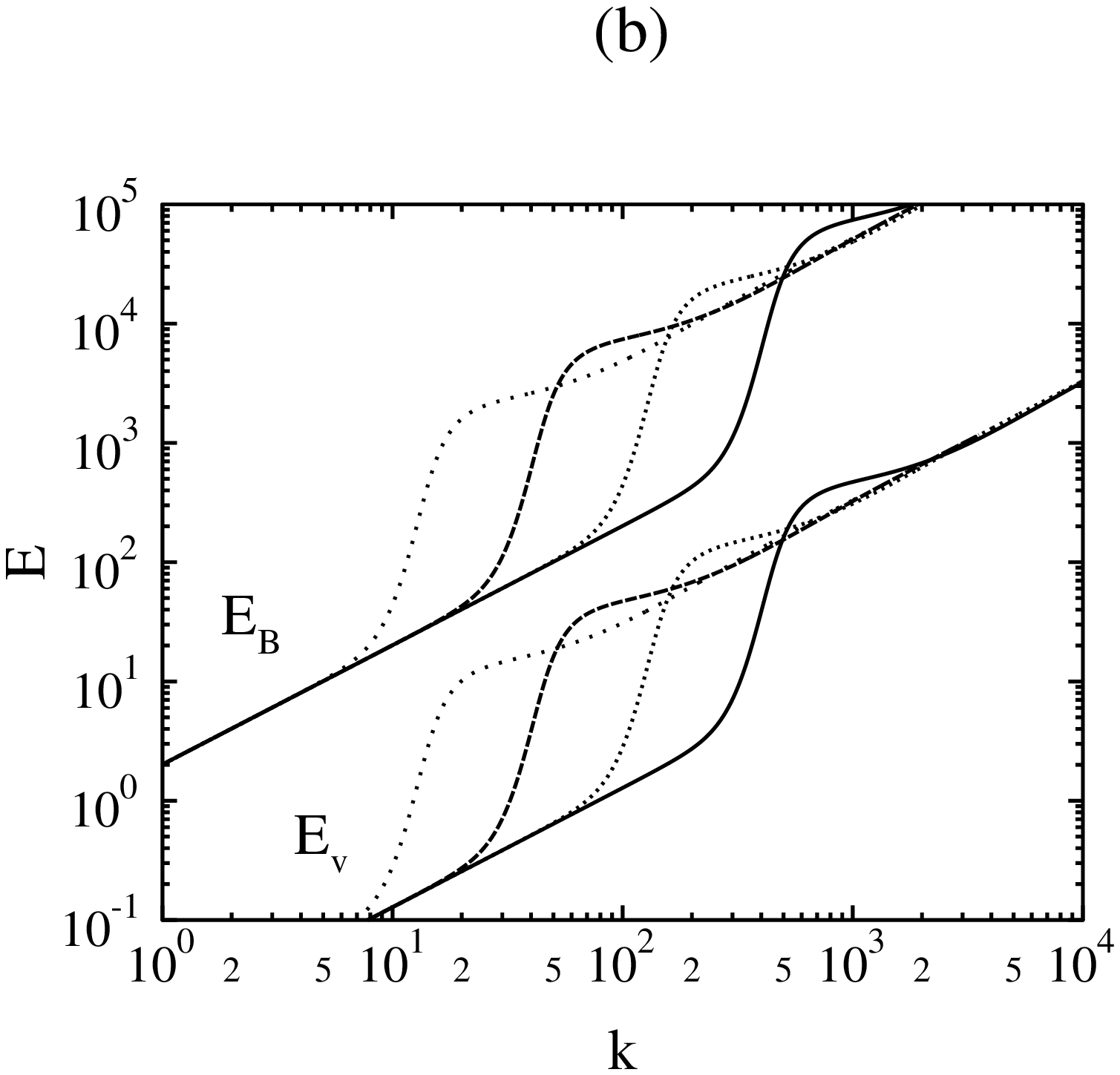}
\caption{The energy spectra in the continuous cascade model for 
(a) small ($\nu=10^{-2}$) plasma viscosity; in this case there is
equipartition and magnetic and kinetic energies are equal; (b) 
large ($\nu=10^{2}$) plasma viscosity; absence of equipartition is
apparent. In both figures the curves
are, from right to left, for conformal times t=1,10,100 and 1000, 
and $\eta=10^{-9}$.
}
\label{kuvadmpw1}
\end{figure}
%\vspace*{40mm}

The scaling relation is in general only valid in the ``inertial range" where 
viscosity can be
ignored, because the viscosity term is inconsistent with the scaling. There is, 
however,
one exception, where scaling and viscosity are consistent, namely $p$=1. Here 
$-u'(x)$
on the left hand side of eq. \rf{u} is replaced by
\be{replace}
-u'(x)-\nu u(x).
\ee
Similarly, in eq.\rf{B} $-B'(x)$ is replaced by
\be{replace2}
-B'(x)-\eta B(x).
\ee   
This is the reason we have preferred to perform the numerical calculations in 
the scaled version of the continuous  cascade model for the case 
$p$=1. We took as the initial conditions $u=m=0,~M=B=1$ and as in the
discrete case, considered two viscosities ($\nu=10^{-2}$ and $\nu=10^{2}$).
The units are again such that initially $\rho=\rho_0=1$.
Magnetic diffusion was set to $\eta=10^{-9}$. The difference with respect 
to the discrete case is the shape of the initial spectrum of $b_n$.
In the discrete case $b_n$ was initially given by $\sqrt{k}$, 
whereas here we took $b_n(k,0)=k$.
The results are depicted
in Fig. 3, where the inverse cascade can  clearly be seen, virtually
independent of viscosity. They also
confirm that equipartition is lost at large $\nu$.

We have also checked the scaling of $b$ for the data presented
in paper 1. This corresponds to $p$=3/2. The scaling variable is thus $x=
k^{5/2}t$, where it is important that $t$ is the conformal time;
in terms of the Hubble time, the scaling variable is $kt_H^{0.2}$. From its
derivation, scaling is only valid in this case in the
inertial range, where viscosity can be ignored. We see that 
within the fluctuations of the raw data, the predictions
are approximately valid for large times (see Fig. 4).

\begin{figure}
\leavevmode
\centering
\vspace*{70mm}
\includegraphics{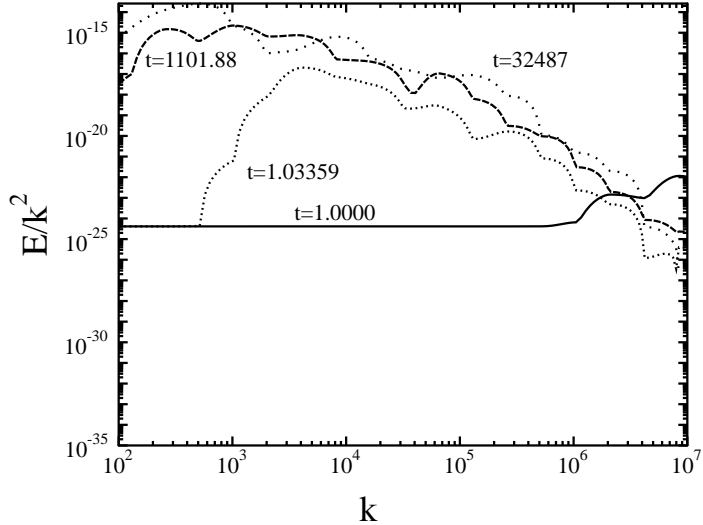}
\caption{The scaled energy spectrum of the discrete cascade model 
at different times (neglecting viscosity).}
\label{kuvascaling}
\end{figure}
%\vspace*{30mm}

Our results very much suggest that in the real MHD, inverse cascade is
operative and is essentially not affected by Silk damping, except
very late and perhaps for very weak fields (which anyhow are not
interesting). 
In our units, in the early universe $\nu\simeq\gg 1$. 
Thus we may conclude that it is unlikely that there is equipartition
in the very early universe. 
Our relativistic approach remains valid roughly until recombination,
after which the plasma becomes matter dominated.
(In the non-relativistic regime $\rho\sim R^{-3}$,
which effectively produces an extra term of the form 
$-(\dot R/R)v$ in the equations of motion. However, because
photon diffusion is so large, this braking due to
expansion is unimportant and conclusions presented above still hold.)
Therefore, it seems plausible that 
small random magnetic domains of the very
early universe may grow to large scale fields,
irrespective of the Silk diffusion.

%%%%%%%%%%%%%%%%%%%%%%%%%%%%%%%%%%%%%%%%%%%%%%%%%%%%%%%%%%%%%%%%
%%%%%%%%%%%%%%%%%%%%%%%%%%%%%%%%%%%%%%%%%%%%%%%%%%%%%%%%%%%%%%%%%%%%%%%%%%%

%\vskip1truecm\noindent
%{\Large\bf Acknowledgements}
%\vskip0.5truecm

\end{document}